\documentclass[manuscript]{aastex}
\usepackage{tikz}
\usepackage{amsmath}

\usepackage{graphicx}

\usepackage{pdflscape}

\slugcomment{\textbf{Revised 2022 January 16} }

\shorttitle{C/2021 A1}
\shortauthors{Jewitt}

\begin{document}

\title{Disintegration of Long-Period Comet C/2021 A1 (Leonard)}


\author{David Jewitt$^{1}$, Yoonyoung Kim$^2$, Michael Mattiazzo$^{3}$, Max Mutchler$^4$,  Jing Li$^1$ \\and Jessica Agarwal$^2$
} 
\affil{$^1$Department of Earth, Planetary and Space Sciences,
UCLA}
\affil{$^2$Institute for Geophysics and Extraterrestrial Physics, TU Braunschweig, D-38106 Braunschweig, Germany}
\affil{$^3$Swan Hill Observatory, Australia}
\affil{$^4$ Space Telescope Science Institute, 3700 San Martin Drive, Baltimore, MD 21218}

\email{jewitt@ucla.edu}

\begin{abstract}

We present imaging observations of the disintegrating long-period comet C/2021 A1 (Leonard).  High resolution observations with Hubble Space Telescope show no evidence for surviving fragments, and place a 3$\sigma$ upper limit to their possible radius $\sim$60 m (albedo 0.1 assumed).    In contrast, wide field observations from the Swan Hill Observatory, Australia, show an extensive debris cloud, the cross-section and estimated mass of which are consistent with complete disintegration of the nucleus near mid- December 2021 (at about 0.8 au).  Two methods give the pre-disruption nucleus radius, $r_n = 0.6\pm0.2$ km.  Tidal, collisional, sublimation and pressure-confined explosion models provide implausible explanations of the disintegration.  However, rotational instability driven by outgassing torques has a very short timescale ($\sim$0.1 year) given the orbit and size of the C/2021 A1 nucleus, and offers the most plausible mechanism for the disruption. Initial rotational breakup is accelerated by the exposure and strong sublimation of previously buried volatiles, leading to catastrophic destruction of the nucleus.

\end{abstract}


\section{INTRODUCTION}
\label{intro}

Comet C/2021 A1 (Leonard), hereafter ``A1'', was discovered on UT 2021 January 3 as a diffuse V $\sim$ 19 magnitude object inbound to the Sun at  heliocentric distance $r_H$ = 5 au  (Leonard 2021).  A1 is a long-period comet, with heliocentric osculating semimajor axis $a$ = -6124 au, eccentricity $e$ = 1.0001 and inclination $i$ = 132.6\degr, reaching perihelion (at $r_H$ = 0.615 au)  on UT 2022 January 03.3, about a year after discovery.  Although presently following a weakly hyperbolic orbit, the pre-entry orbital elements (corrected for planetary perturbations to 1900 January 1, when the heliocentric distance was 137 au) are those of a bound object, $a$ = 2020 au, $e$ = 0.999696 and $i$ = 132.7\degr.  A1 is thus not a dynamically new comet, having passed through the planetary system $\sim10^5$ years ago.

Comet A1 attained naked eye visibility in late 2021 and then displayed spectacular gas and dust tails.  However, images and commentary recorded in public on-line archives\footnote{e.g.~\url{https://britastro.org/cometobs/2021a1/thumbnails.html}} indicate that A1 became photometrically unstable in 2021 December and 2022 January.  Measurements of the OH production rate from the Nancay radio telescope were steady near $Q_{OH}$ = 2.6$\times10^{28}$ s$^{-1}$ between UT 2021 December 9 and 12, but jumped by a factor of $\sim$8 to $Q_{OH}$ = 22$\times10^{28}$ s$^{-1}$ on December 15, even as the heliocentric distance barely decreased from 0.80 au to 0.74 au \citep{Crovisier21}.   The morphology also changed, becoming more diffuse and with ``the tail being more prominent than the head'' on UT 2022 January 22\footnote{\url{https://groups.io/g/comets-ml/message/30541}}  at $r_H \sim$ 0.74 au outbound.   Based on these early observational reports we requested Director's Discretionary Time on the Hubble Space Telescope (HST), with the science objective being to study the presumed breakup of this long-period comet at the highest angular resolution.   Independently, coauthor Mattiazzo also obtained  wide-field imaging data using a private  telescope at the Swan Hill Observatory in Australia.    The wide-field and HST data are  highly complementary, with the former providing sensitivity to low surface brightness debris over a wide angle and the latter providing ultra-high resolution and very deep imaging of the near-nucleus region. 

While the phenomenon of cometary breakup has been known for over a century, very few physical observations of disintegrating comets are to be found in the refereed literature.  In this paper, we present the observations  and consider possible causes of the breakup of comet A1.

\section{OBSERVATIONS}
\subsection{Hubble Space Telescope}
The 2.4 m diameter Hubble Space Telescope was used to observe disintegrating A1 under program GO 16929.  We used the WFC3 camera, which houses two 2015$\times$4096 pixel charge coupled devices separated by a 1.2\arcsec~wide gap.  The   0.04\arcsec~pixel$^{-1}$ image scale gives a full-frame 162\arcsec$\times$162\arcsec~field of view.   HST images were taken using the F350 LP filter in order to maximize throughput.  This filter has an effective central wavelength $\lambda_c$ = 6230\AA~when observing a Sun-like (G2V) source and a FWHM $\Delta \lambda$ = 4758\AA.    We secured four images each of 450 s duration in each of the first three orbits and five frames of 285 s, with a sub-frame readout, in the fourth.  The first three orbits were obtained in 2022 April with spacings of one and four days, with the intention being to measure the sky-plane motions of fragments produced by the break-up of A1.  The fourth orbit was scheduled on UT 2022 June 7 to coincide with the passage of the Earth  through the projected orbit plane of the comet.  Observations from this vantage point provide a model-free measure of the thickness of the dust distribution perpendicular to the plane.   Unfortunately, the images from the fourth orbit suffered from extreme field star contamination, as a result of the low (-6\degr) galactic latitude of the comet, and were not useful.

\subsection{Swan Hill Observatory}
Wide-field observations were taken by co-author Michael Mattiazzo using a 0.28 m diameter, f/2.2 wide-field telescope at the Swan Hill Observatory (observatory code Q38), located in Victoria, Australia.   A 4655$\times$3522 pixel CMOS imaging device (Panasonic model QHY163M) provided an image scale of 1.27\arcsec~pixel$^{-1}$, and a field of view  approximately 1.6\degr$\times$1.2\degr.   Each pixel of the 0.28 m telescope subtends a solid angle equal to 10$^3$ HST pixels.  Ten images each of 30 s duration were obtained, during which time the comet moved relative to field stars by about 2.7\arcsec, which is small compared to the  5.1\arcsec~full width at half maximum of point source objects in the data.   The wide field image shows evidence for loss of sensitivity due to vignetting, especially near the corners of the device.  We removed this by fitting a cubic spline surface to the image, using the median signal within 50$\times$50 pixel boxes (after checking that the procedure did not self-subtract the comet).  

No filter was employed in order to maximize the throughput of the system.  The quantum efficiency of the detector peaks near a central wavelength  5500\AA, and has a FWHM   estimated at $\sim$4000\AA.  The central wavelength is close to that of Johnson V (see the discussion in Bessel 1990), but the  response is so broad that it captures the same light as the Johnson B, V and R filters (or, equivalently, the Sloan g and r filters) combined.  The large bandwidth and lack of a standard filter together limit the accuracy with which the measured magnitudes can be related to, for example, the V band magnitudes.   We  calibrated the data using measurements of field stars on the Sloan filter system, provided by the Skymapper southern survey (Wolf et al.~2018).  For this purpose we extracted measurements using circular apertures of projected radius 12.7\arcsec, with sky subtraction from the median signal within a concentric annulus having inner and outer radii 19.1\arcsec~and 38.1\arcsec, respectively. In order to minimize the color term in our photometry, we selected stars with optical color g-r $\sim$0.4 to 0.5, so as to approximately match the color of the Sun (given as g-r = 0.45$\pm$0.02 by Holmberg et al.~2006).  We further selected these stars to lie within $\sim$1\arcmin~of the comet in order to minimize spatial variations in the photometry caused by imperfect flatness of the data.

The geometrical circumstances of observation are given in Table \ref{geometry}.

\section{RESULTS}

\subsection{High Resolution Data}
\begin{figure}
\epsscale{0.99}
\plotone{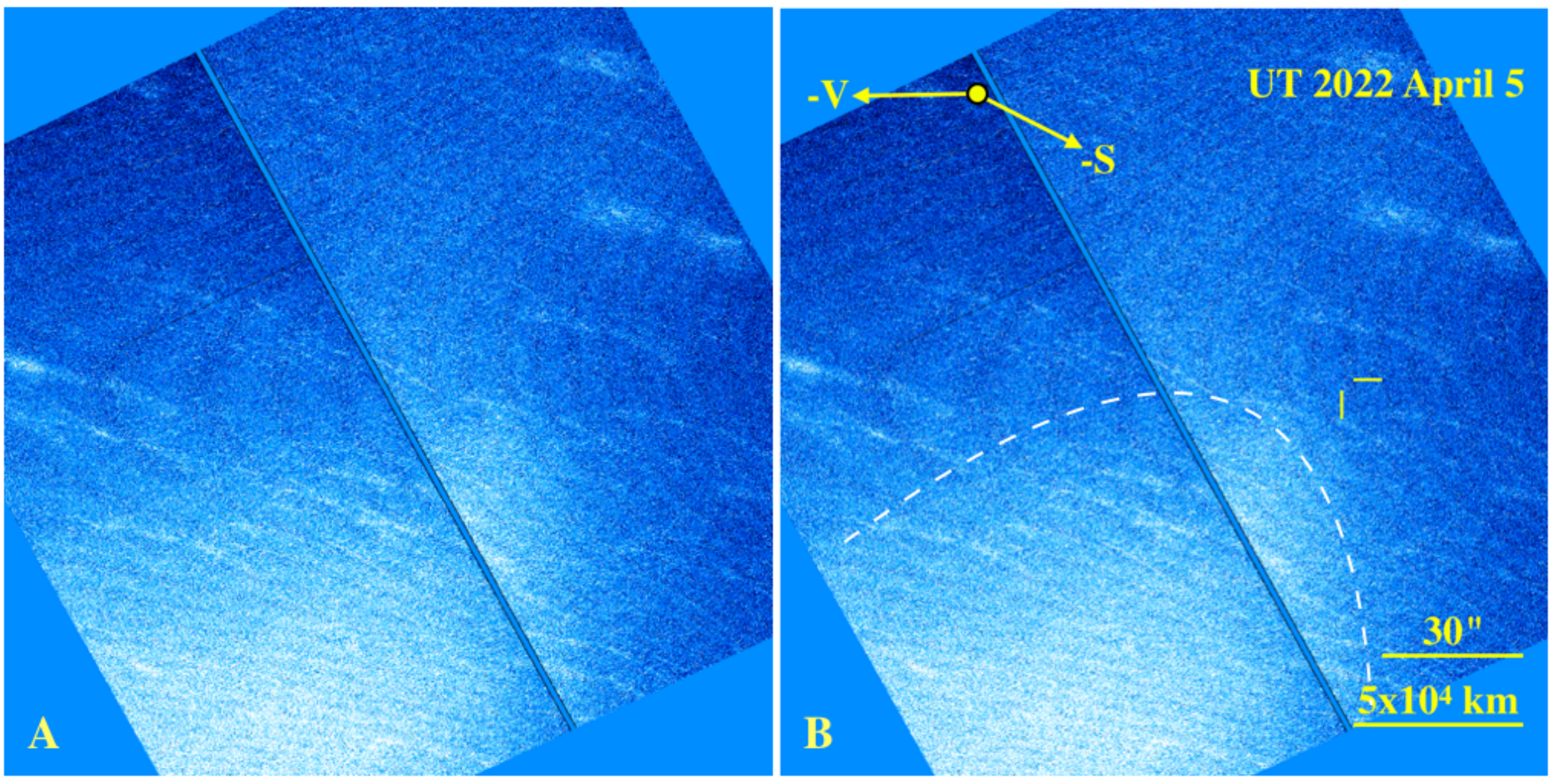}
\caption{A) Composite of four, 450 s HST images from UT 2022 April 5.  Diffuse streaks are imperfectly removed field stars and galaxies.  B) Same image, anotated to show the approximate boundary of the debris (white dashed line) and the  expected location of the nucleus (yellow line segments).  Two scale bars of 30\arcsec~and 5$\times10^4$ km in length are shown, as well as the projected anti-solar ($-S$) and negative heliocentric velocity ($-V$) vectors. North is to the top, East to the Left.   \label{hst_apr5}}
\end{figure}

We combined the four images from each orbit in order to reject cosmic rays, suppress trailed field objects, and reach a fainter limiting magnitude.  The composite from UT 2022 April 5 is shown in  Figure \ref{hst_apr5}; composites from April 6 and 10 look the same.  The predicted location of the nucleus is indicated in the Figure.  The JPL Horizons ephemeris for April 5 gives 3$\sigma$ positional uncertainties of $\pm$1.3\arcsec~in right ascension and $\pm$1.0\arcsec~in declination, both of which are negligible compared to the 160\arcsec~field of view of WFC3.  We searched for the principal nucleus and discrete fragments in the data by comparing image subsets to identify correlated motion, but found none.    Instead, the images  show  evidence for diffuse light scattered from cometary dust, evident in Figure \ref{hst_apr5} as a region of slightly higher surface brightness in the south east quadrant of the image (marked by a dashed white line in the right-hand panel of the figure).  Although it at first resembles a flat-field defect or a smudge of internally scattered light, two lines of evidence show that this region of diffuse brightness is neither. First, the enhanced region is fixed with respect to the daily predicted ephemeris position of A1.  Second, the enhanced region moves on the detector as the telescope orientation angle changes. The enhancement appears at the same position in image composites from all three dates in April, whereas scattered light from bright stars outside the WFC3 field of view would vary as the background stars are completely different from day to day.   A flat-field defect would not rotate as the telescope orientation changes.  We conclude that the diffuse light is sunlight scattered  from cometary debris released from the now invisible nucleus of A1.

The on-line WFC3 Exposure Time Calculator\footnote{\url{https://etc.stsci.edu/etc/input/wfc3uvis/imaging/}} gives a 3$\sigma$ limit for detection of point source objects at V = 26.7, in each of our orbits.  This limiting magnitude is consistent with the measured sky noise in the data.  Corrected to absolute magnitude using phase coefficient $\beta$ = 0.04 magnitude degree$^{-1}$, we find $H \ge$ 22.81.  For a nominal albedo, $p_V$ = 0.1, this corresponds to a 3$\sigma$ limit to the fragment radius, $r \le$ 60 m.

\subsection{Wide Field Data}
\begin{figure}
\epsscale{0.95}
\plotone{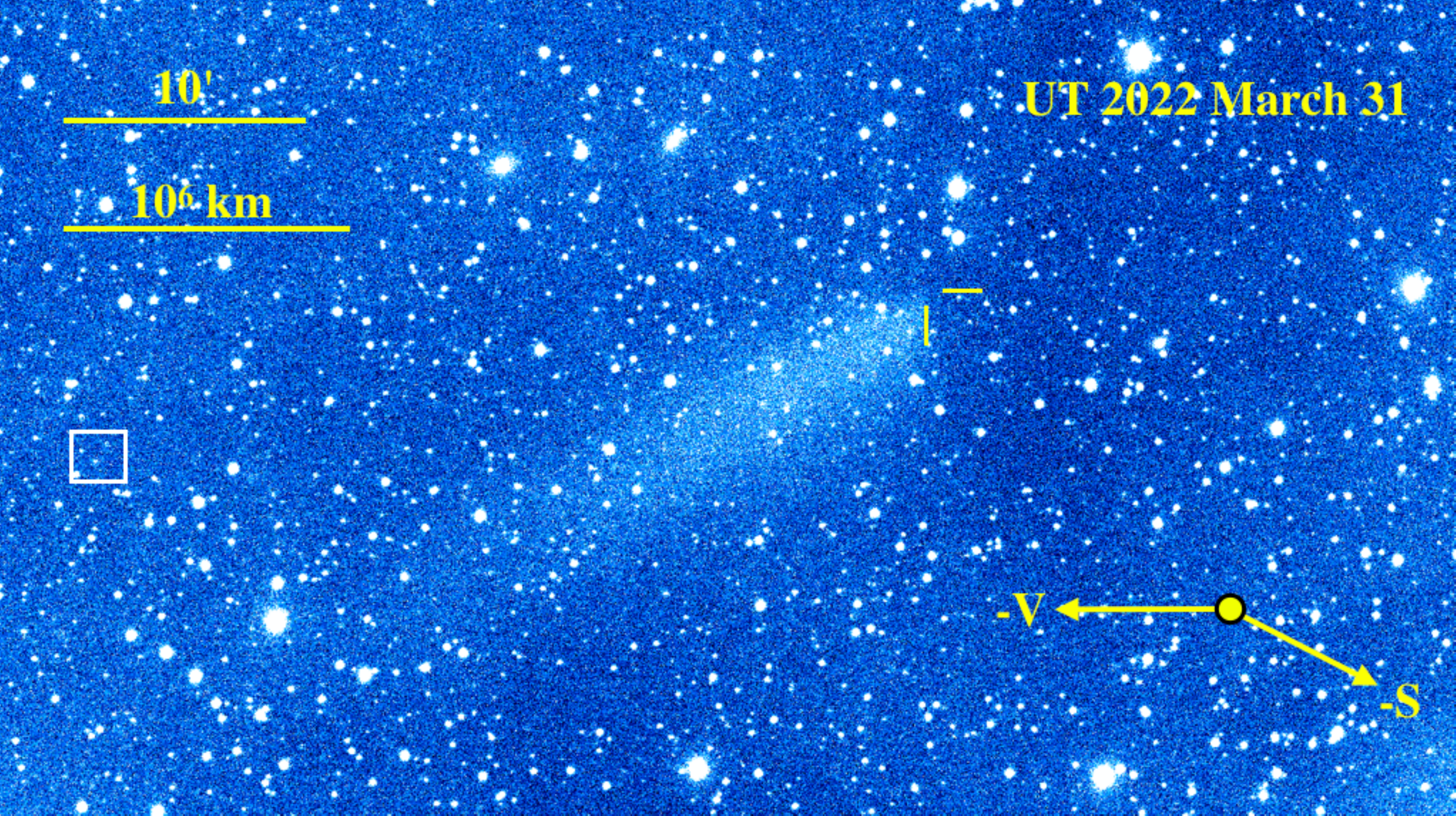}
\caption{Wide field image from Swan Hill Observatory showing C/2021 A1 on UT 2022 March 31.  10\arcmin~and 10$^6$ km scale bars are shown, as well as the projected anti-solar ($-S$) and negative heliocentric velocity ($-V$) vectors. Yellow lines mark the ephemeris location of the nucleus. The white square shows the size of the HST field of view.  The image has North to the top, East to the left.  \label{wideimage}}
\end{figure}

The composite wide field image is shown in Figure \ref{wideimage}. A low surface brightness dust structure extends over at least 0.4\degr~(2$\times10^6$ km in the plane of the sky), with a position angle 120\degr$\pm$2\degr~and no  indication of a brightness peak at the expected location of the nucleus.   The latter was determined from the JPL Horizons ephemeris for the mid-time of the image, and is marked in the figure.  Overall, the morphology is similar to that of C/2010 X1 (Elenin), a long period comet which disintegrated when inbound near $r_H$ = 0.6 AU (Li and Jewitt 2015), and C/2019 J2 (Palomar), which disintegrated pre-perihelion near $r_H$ = 1.9 au (Jewitt and Luu 2019). Comparison with Figure \ref{hst_apr5} shows that the HST, which was pointed at the expected location of the nucleus, indeed recorded diffuse light from the western tip of this dust structure.

\begin{figure}
\epsscale{0.75}
\plotone{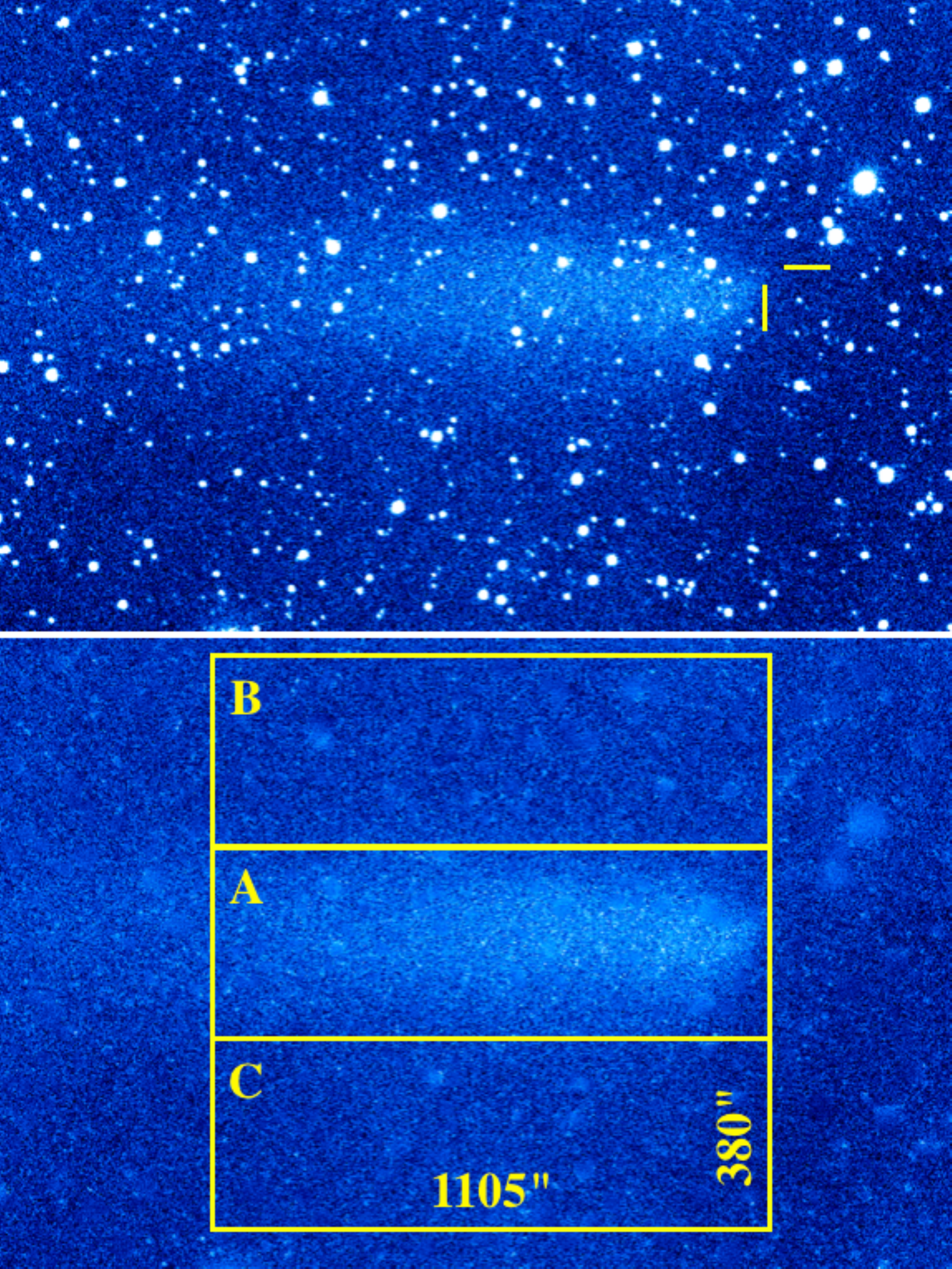}
\caption{(Upper:) Same image as in Figure \ref{wideimage} but rotated to bring the axis of the dust tail to the horizontal and shown at a larger scale.  Yellow lines mark the ephemeris location of the nucleus.  (Lower:)  Locations of the photometry regions A, B and C used to measure the scattering cross-section of particles in the tail. \label{boxes}}
\end{figure}

We estimated the total light from the dust as follows.  First, we rotated the image to bring the long axis of the dust tail to the horizontal (upper panel in Figure \ref{boxes}).  Next, we manually replaced field stars  with the average of surrounding pixels.  The median signal from the comet was then computed within a rectangular box, ``A'' in the lower panel of Figure \ref{boxes}) 1105\arcsec~long by 380\arcsec~tall, and the background sky estimated from equal-sized photometry boxes contiguous with the comet box but displaced above and below it (``B'' and ``C'' in Figure \ref{boxes}).  Figure \ref{boxes} shows that the tail extends beyond the left edge of the photometry box ``A'' but the increased uncertainty imposed by the sky rendered measurements of this very faint material impractical.   The light from the tail was calculated from $f_T = f_A - (f_B + f_C)/2$, where $f_x$ is the flux in box ``x''.  Then, applying the calibration obtained from field stars, we  find $V_T = 10.9\pm$0.5, where the quoted error is our best estimate of the uncertainty resulting from non-flatness of the data, the transformation from the wide response of the camera and the effective V magnitude.   With assumed phase function 0.02$\pm$0.02 magnitude degree$^{-1}$ and the geometry given in Table \ref{geometry}, the corresponding absolute magnitude is $H$ = 7.6$\pm$0.6, where the larger uncertainty is introduced by the phase correction.   The scattering cross-section needed to give this absolute magnitude is $C = 1.4_{-0.8}^{+1.0}\times10^{10}$ m$^2$, assuming geometric albedo $p_V$ = 0.1 (appropriate for cometary dust; Zubko et al.~2017). 

\begin{figure}
\epsscale{0.95}
\plotone{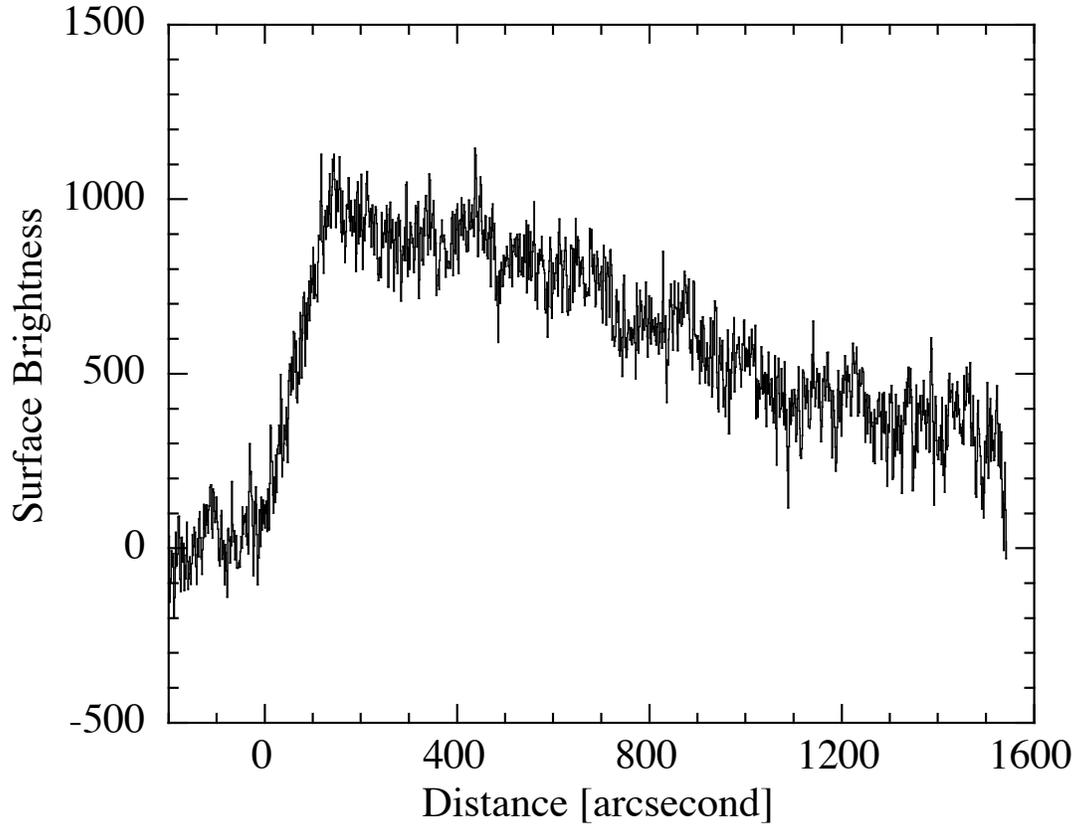}
\caption{Surface brightness profile parallel to the long axis of Box A (Figure \ref{boxes}) plotted against the distance from the nucleus ephemeris location (axis is reversed relative to Figure \ref{boxes}).  1000 units  correspond to a surface brightness $\Sigma$ = 24.4 magnitudes arcsec$^{-2}$.  The linear distance scale is approximately 1500 km per arcsecond.  \label{parallel}}
\end{figure}

Figure \ref{parallel} shows the averaged surface brightness profile from the March 31 image, measured parallel to the long axis of region A in Figure \ref{boxes}.  Most of the scatter in the surface brightness profile is statistical noise in the data, but larger oscillations (for example at $\sim$480\arcsec~and 750\arcsec)  result from spatial background variations caused by the digital removal of field stars.   In this plot, the peak  of 1000 units corresponds to a surface brightness $\Sigma$ = 24.4 magnitudes arcsec$^{-1}$, about 5\% of the surface brightness of the night sky.  The
 surface brightness shows a steep increase, reaching a maximum at about 100\arcsec ~from the ephemeris nucleus location, followed by a steady decline at larger projected angles.    This profile shape is indicative of a suddenly terminated dust mass release, with the peak of the profile giving the distance traveled by the largest, slowest particles.   
\clearpage

\section{DISCUSSION}

\subsection{Radius and Mass of the Nucleus}
\label{nucleus}

We use the effective spherical nucleus radius  of A1 $\bar{r}$ = 0.6$\pm$0.2 km from Jewitt (2022).  This estimate is based on independent measurements of $Q_{H_2O}(1)$, the gas production rate at 1 au, and of $\alpha_1$, the non-gravitational acceleration at 1 au.  Comet A1 has $Q_{H_2O}(1)$  = 1.9$\times10^{28}$ s$^{-1}$ (only pre-perihelion observations are used because post-perihelion rates are clearly affected by the breakup) and $\alpha_1$ = 1.3$\times10^{-6}$ m s$^{-2}$, provided by JPL Horizons.  A substantially smaller nucleus would have a surface area insufficient to supply the $Q_{H_2O}(1)$, while a  substantially larger nucleus would have too much mass to be accelerated at $\alpha_1$ given the known gas production rate.  
Using $\bar{r}$ and nominal nucleus density $\rho_n$ = 500 kg m$^{-3}$ \citep{Groussin19}, we estimate the nucleus mass $M_n = (4.5_{-3.2}^{+6.5})\times10^{11}$ kg. The largest surviving fragments, with radii $<$60 m, individually contain $<10^{-3}$ of the mass of the primary.

\subsection{Time of Disruption}
\begin{figure}
\epsscale{0.95}
\plotone{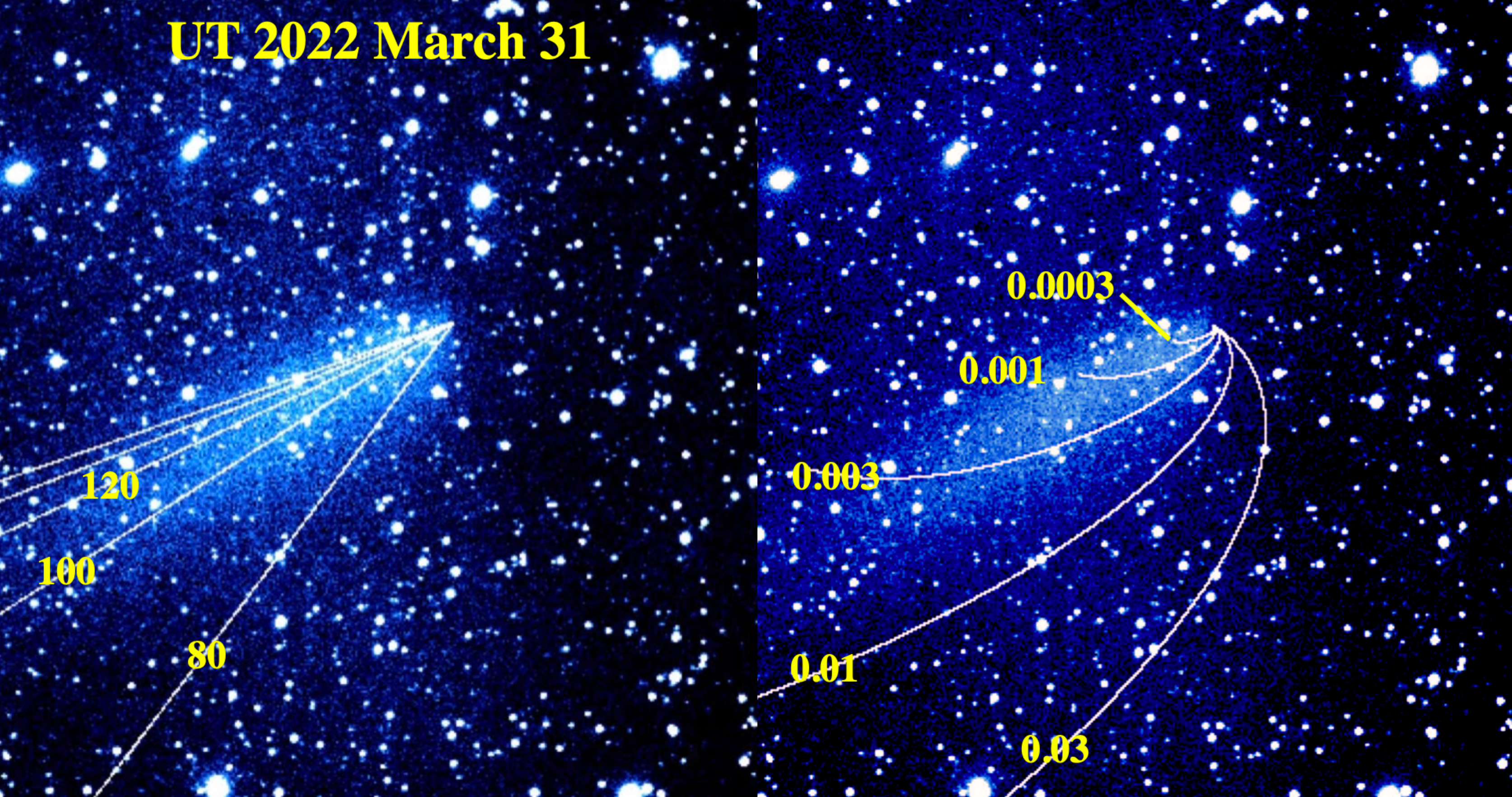}

\caption{(Left:) Same image as Figure \ref{wideimage} with synchrones overplotted, for ejection dates 80, 100, 120, 140 and 160 days prior to the date of the image.    (Right:) Syndynes for particles with $\beta$ = 0.0003, 0.001, 0.003, 0.01 and 0.03, as marked.  The axis of the debris cloud is best matched by the  110$\pm$10 day synchrones, corresponding to ejection on UT 2021 December 11$\pm$10.  \label{synchrones}}
\end{figure}

Syndynes (the loci of particles having one size, released with zero initial relative velocity over a range of times; \cite{Finson68}) are curved and do not match the linear shape of the debris cloud in A1.  Instead, the  morphology more resembles a set of synchrones as shown in Figure \ref{synchrones}.   Synchrones trace the loci of particles in the sky plane having a range of sizes (hence, radiation pressure accelerations) but released from the nucleus simultaneously.  The position angle of the debris trail in A1 is most compatible with ejection 110$\pm$10 days before the image was taken, i.e.~on UT 2021 December 11$\pm$10. This is about a month before reports of distinct morphological change appeared but coincides with a dramatic increase in the OH production rate from 4.4$\times10^{28}$ s$^{-1}$ on UT 2021 December 19 to 14$\times10^{28}$ s$^{-1}$ on UT 2021 December 21,  in unpublished SOHO/SWAN data (personal communication M. Combi).  It is also close to a reported OH outburst on UT 2021 December 15 \citep{Crovisier21}.  While we lack continuous coverage of the gas production from A1, it is likely that the sublimation rate became highly unstable as a result of the breakup of the nucleus when close to perihelion.

We assume that the disintegration began on UT 2021 December 11$\pm$10.  To reach the far end of the measured debris cloud (an angular distance $\sim$1500\arcsec, corresponding to linear distance $L = 2.2\times10^6$ km) under the action of radiation pressure requires an average acceleration $2L/\Delta T^2$, where $\Delta T$ = 111 days (9.6$\times10^6$ s) is the interval between the time of disintegration and the Swan Hill image from UT 2022 March 31.  In units of the solar gravitational acceleration at the average $r_H$ = 1.3 au heliocentric distance in this period, 

\begin{equation}
\beta = \frac{2L r_H^2}{g_{\odot}(1) \Delta T^2}
\label{beta}
\end{equation}


\noindent where $g_{\odot}(1)$ = 0.006 m s$^{-2}$ is the solar gravity at 1 au and $r_H$ is expressed in au.  Substituting, we obtain $\beta$ = 0.01.  With $\beta \sim 1/a_{\mu m}$, where $a_{\mu m}$ is the particle radius expressed in microns (c.f. \cite{Bohren83}), we infer that the particles at the far end of the tail in the March 31 image had $a_{\mu m} \sim$ 75 $\mu m$.   All particles in the visible debris cloud on UT 2022 March 31 must be larger, while smaller particles were  presumably ejected but have been swept by radiation pressure beyond the visible extent of the tail.   Particles near the peak of the surface brightness profile (angular distance $\sim$100\arcsec, corresponding to $L = 1.4\times10^5$ km) have $\beta \sim 10^{-3}$ by Equation \ref{beta} and, therefore, radii $\sim$1 mm. 

\subsection{Mass of the Optical Debris}
\label{debris_mass}
How does the mass of the debris compare with the mass of the nucleus prior to its disappearance?  To answer this question, we treat the debris as consisting of a distribution of spherical particles with radii between $a$ and $a + da$ written as $n(a)da$.  Then, the combined mass of the particles between minimum radius $a_1$ and maximum radius $a_2$ is

\begin{equation}
M_d = \int_{a_1}^{a_2} \frac{4}{3} \pi \rho a^3 n(a) da
\label{M}
\end{equation}

\noindent while their combined cross-section is 

\begin{equation}
C = \int_{a_1}^{a_2} \pi a^2 n(a) da
\label{C}
\end{equation}

\noindent It is useful to represent the size distribution as a power law

\begin{equation}
n(a) da = \Gamma a^{-\gamma} da
\label{distribution}
\end{equation}

\noindent where $\gamma$ is the differential size distribution index and $\Gamma$ is a normalizing constant.  Substituting equation \ref{distribution} into equations \ref{M} and \ref{C} and eliminating $\Gamma$, we obtain

\begin{equation}
M_d = \frac{4}{3} \rho C \frac{\int_{a_1}^{a_2}  a^{3-\gamma} da}{\int_{a_1}^{a_2} a^{2-\gamma} da} 
\label{MoverC}
\end{equation}

\begin{figure}
\epsscale{0.7}
\plotone{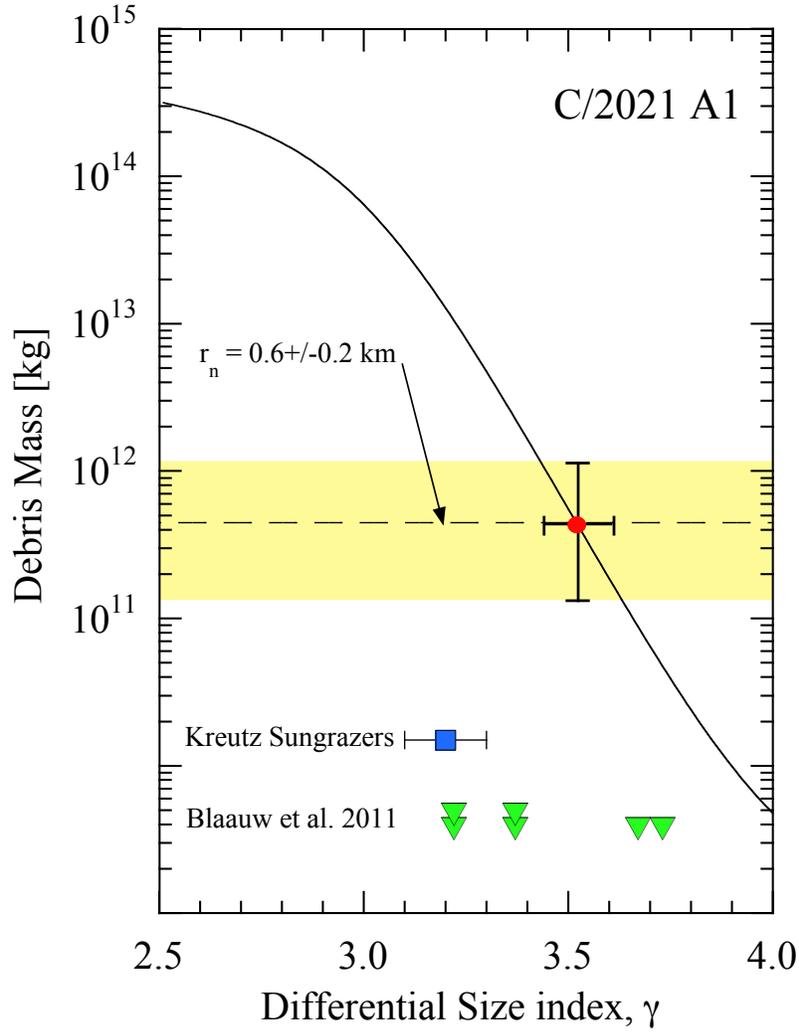}
\caption{Total mass of the debris cloud (assuming density $\rho_n$ = 500 kg m$^{-3}$) plotted as a function of the differential power law index, $\gamma$, is plotted as a solid black line.  The equivalent spherical mass of the original 0.6$\pm$0.2 km radius nucleus is shown (assuming the same $\rho_n$), together with its uncertainty, as a yellow horizontal band.  The debris and nucleus masses are equal at $\gamma = 3.5\pm0.1$, shown by the red filled circle.  For comparison we show, as a blue square, the size distribution of the Kreutz sungrazing comets \citep{Knight10} and, as green triangles, several radar-measured meteoroid streams \citep{Blaauw11}.  The vertical positions of the Kreutz and radar stream points have no meaning. \label{mass_plot}}
\end{figure}

The minimum particle radius is selected as $a_1$ = 75 $\mu$m, since all smaller particles would have been swept out of the image field in the time since ejection.   The maximum radius, $a_2$ = 60 m, is set by the non-detection of larger bodies in our deep HST imaging data. With these values for $a_1$ and $a_2$, we plot Equation \ref{MoverC}  as a function of $\gamma$ in the range 2.5 $\le \gamma \le$ 4.0  (Figure \ref{mass_plot}).   The particle mass required to account for the measured cross-section, $C$, is seen to vary by orders of magnitude for modest changes in the index, $\gamma$, with smaller values (flatter distributions) hiding a larger fraction of the total mass in big bodies.  

Also plotted in the figure is the nucleus mass, $M_n = (4.5_{-3.2}^{+6.5})\times 10^{11}$ kg, computed from the effective radius, $r_n$ = 0.6$\pm$0.2 km, (Section \ref{nucleus}), and density, $\rho_n$ = 500 kg m$^{-3}$, with the mass uncertainty marked as a horizontal yellow band.  The red point marks the intersection of the two curves where $M_d = M_n$ and shows that, for index $\gamma =$ 3.5$\pm$0.1, the debris mass and nucleus mass are equal.   The upper limit to the size distribution could be substantially smaller than the 0.6 km limit set by the Hubble data, in which case a smaller value of the index would be needed for the mass of the debris to equal the mass of the nucleus.  A relevant comparison can be made with the  size distribution of the Kreutz sungrazing comets, which are themselves produced by the fragmentation of a precursor body.  The Kreutz objects have $\gamma$ = 3.2 in the 5 m to 35 m radius range \citep{Knight10}, plotted as a blue square in Figure \ref{mass_plot}.  The uncertainty on $\gamma$ for the Kreutz objects is not stated; we have plotted a nominal $\pm$0.1 error bar for reference and note  reasonable agreement with the  index deduced for A1 within the uncertainties.
Perhaps less relevant are radar measurements of the debris size distributions in six meteoroid streams, most associated with decaying comets. These are plotted for comparison using green triangular symbols (\cite{Blaauw11}).  The formal meteoroid stream index uncertainties are comparable to the size of the symbols in the figure.  The measured indices span the range $\gamma$ = 3.2 to 3.7, encompassing the values found for A1 and the Kreutz comets.   

We conclude that the optical cross-section presented by the debris in 2022 March is consistent with the complete disintegration of the original $\sim$0.6 km scale nucleus into a power law distribution (index $\gamma =$ 3.5$\pm$0.1) of particle sizes.  We emphasize that we possess no independent evidence that the debris mass and original nucleus mass are equal, although a consideration of the particle properties using more detailed considerations (section \ref{MC}) supports this result.  It should also be noted that 60 m is an upper limit to the size of the largest post-disruption ``particles'' and our result would be changed if $a_2 \ll$ 60 m, as it would if the size distribution of particles is not well represented by a single power law across the full range of sizes.  It is also not obvious that the density of the particles should necessarily be the same as the bulk density of the nucleus, as we have assumed.  These and other physically plausible possibilities lie beyond the observational constraints obtained from the data.

\subsection{Monte Carlo Simulation}
\label{MC}
We next used a Monte Carlo simulation as developed by \cite{Ishiguro07} (see also \cite{Kim17}) to model the cometary debris in more detail. The model is under-constrained  and cannot provide unique solutions for the particle properties. It is nevertheless valuable in allowing us to test the deductions made based on order of magnitude considerations, and also to more fully explore the range of plausible solutions.   We particularly examined the effect of the particle size distribution index and the minimum and maximum particles sizes in the distribution.   

\begin{figure}
\epsscale{0.95}
\plotone{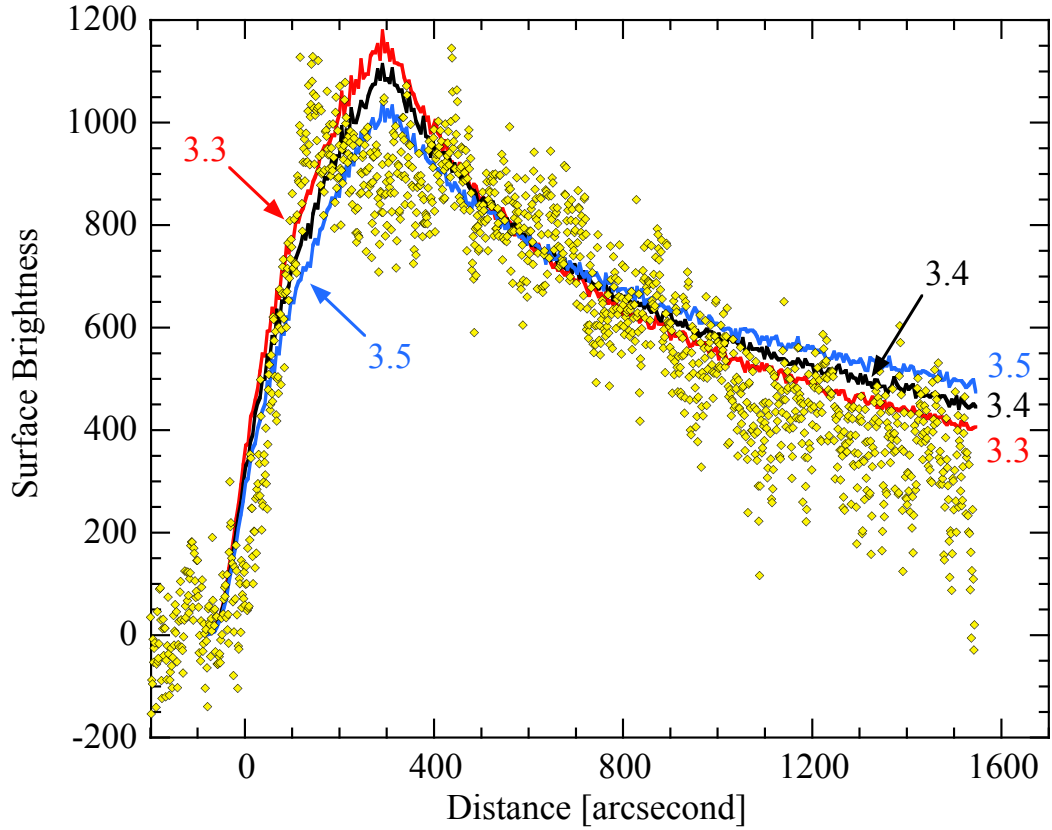}

\caption{Axial surface brightness profile on UT 2022 March 31 (yellow diamonds) compared with results from a Monte Carlo simulation.  The models shown have size index $\gamma$ = 3.3 (red curve), 3.4 (black curve) and 3.5 (blue curve), all with  $7\times10^{-4} \le \beta \le$ 0.07, corresponding to particle radii 14 $\mu$m to 1.4 mm.  \label{sb_models}}
\end{figure}

Figure \ref{sb_models} shows the data with results of simulations for $\gamma$ = 3.3, 3.4 and 3.5 and size parameter in the range $7\times10^{-4} \le \beta \le$ 0.07, with ejection on 2021 December 11.  The upper limit to $\beta$ (lower limit to particle radius)  is set by the field of view, with smaller particles have already been pushed out of the field by radiation pressure.  We obtain $a \ge$ 14 $\mu$m, different by a factor of five from the limit $a \ge$ 75 $\mu$m estimated by the order of magnitude procedure, above.  The lower limit to $\beta$ (upper limit to the particle size of $\sim$1.4 mm) is determined from the location of the surface brightness peak in Figure \ref{sb_models}.  This is very small compared to the  60 m upper limit to the radius of the largest possible fragment, set by non-detection  in the HST images.  However, this difference is understandable since, for commonly measured cometary size distributions, the scattering cross-section is dominated by the smallest particles; large particles contribute little to the cross-section and thus are poorly constrained by scattered light observations.  In order to fit the data, we assumed that the particle ejection speed varies with size parameter as $V = V_0 \beta^{1/2}$, with $V_0$ = 550 m s$^{-1}$ being the gas thermal speed.  Unlike the particle trails of weakly active comets and asteroids, a high ejection speed  is required in order to fit the large width of the debris cloud in A1.  

As is evident in Figure \ref{sb_models}, the plotted models do not perfectly reproduce the measured surface brightness profile, with larger $\gamma$ models being 25\% to 30\%  brighter than the data  at large distances from the nucleus and smaller $\gamma$ models being too sharply peaked compared to the measurements.  If they are real, these differences could result from physical effects not included in the model.  For example, we have ignored dust released before disintegration, reasoning that the dramatic outbursts and brightening starting in mid-December would swamp any signal from older material.  As another example, large aggregate grains in the tail might break up into smaller particles which would be quickly swept from the field of view by radiation pressure, perhaps explaining the lower brightness of the tail $\gtrsim$1000\arcsec~from the nucleus.  On the other hand, the differences between the models and the measured profile are certainly affected by systematic uncertainties intrinsic to the wide field data, particularly by imperfect flatness of the data and by the presence of scattered light from bright background sources.    Rather than over-interpret the data, we conclude from the Monte Carlo simulation only that $\gamma \sim 3.4\pm0.1$ provides a broad match to the profile, while much steeper and much less steep distributions do not.  The range of allowable indices deduced from Monte Carlo models is consistent with  $\gamma = 3.5\pm0.1$ as inferred from the debris mass in Section \ref{debris_mass} (c.f.~Figure \ref{mass_plot}).

Lastly, we used the Monte Carlo model to test the possibility that the debris observed in 2022 March  could be long-lived material released before perihelion, in the form of a so-called ``neck-line'' structure (e.g.~Pansecchi and Fulle 1990).  We find that material ejected in the period 2021 November 15 to December 15 would produce a tail structure in March having position angle (113\degr) distinctly different from that measured (120\degr) or calculated from the impulsive ejection model (119\degr).   In addition, neck-line structures in other comets are most prominent when observed from near the projected orbital plane, whereas our observations were taken $\sim$20\degr~from the orbital plane of C/2021 A1 (c.f.~Table \ref{geometry}).  The combination of the unfavorable observing geometry, the failure to reproduce the measured position angle of the dust in 2022 March, and the obvious importance of the outbursts reported in 2021 December together show that pre-perihelion dust is a negligible contributor to the post-perihelion appearance.

\subsection{Disintegration Mechanism}
The preceding discussion shows that a $\sim$0.6 km scale nucleus disintegrated into fragments, the largest of which were no more than about 10\% of the radius of the original body.  What process could lead to such a dramatic outcome?

\textbf{Tidal Breakup:}   The 0.615 au perihelion distance of A1 far exceeds the Roche radius of the Sun ($\sim$10$^{-2}$ au), negating the possibility of a tidal breakup.  Comet A1 did pass within a distance $r_V$ = 0.029 au from Venus on UT 2021 December 18   \citep{Zhang21} but this is still $\sim$300 times the Roche radius ($\sim$10$^{-4}$ au) of the planet.  To within a numerical multiplier, the differential of the gravitational force on opposite sides of the nucleus is $\Delta F \sim (G M_V \rho_n r_n^3/r_V^2) (r_n/r_V)$ giving an order of magnitude tidal stress  $S \sim \Delta F/r_n^2$ or

\begin{equation}
S \sim \frac{G M_V \rho_n r_n^2}{r_V^3},
\end{equation}

\noindent where $G = 6.67\times10^{-11}$ N kg$^{-2}$ m$^2$ is the gravitational constant, $M_V = 5\times10^{24}$ kg is the mass of Venus and the other quantities are already defined.  Substituting  $\rho_n$ = 500 kg m$^{-3}$, $r_n$ = 600 m, and $r_V$ = 0.029 au, we estimate $S \sim 10^{-6}$ N m$^{-2}$ at closest approach, which is orders of magnitude smaller even than the cohesive strengths of fine, unconfined powders ($S \gtrsim$ 100 N m$^{-2}$) measured in the laboratory (Garcia-Trinanes et al.~2019).  The disintegration of A1 is very unlikely to be a consequence of tidally induced stresses.

\textbf{Equilibrium Sublimation:} The  rate of loss of surface material is $dr_n/dt \sim -f_s/\rho$, where $f_s \sim 2\times10^{-4}$ kg m$^{-2}$ s$^{-1}$, at 1 au. Substitution gives $dr_n/dt \sim$ -3 mm day$^{-1}$.  At this rate, the timescale for eroding the whole nucleus would be $|r_n/(dr_n/dt)| \sim$ 40 years, which is very large compared to the $\sim$1 year spent by A1 in the vicinity of the Sun. In any case, sublimation would produce steady erosion of the comet not a catastrophic disintegration like that observed.  Equilibrium sublimation  cannot account for the sudden disintegration of A1.

\textbf{Collisional Disruption:} Collisional disruption timescales for 0.6 km scale objects, even in the dense parts of the asteroid belt, are measured in hundreds of millions of years \citep{Bottke05}.  Comet A1 arrived from a high inclination orbit and disintegrated $\sim$0.5 au from the ecliptic plane where there are no known objects with which to collide.  We confidently dismiss the possibility that A1 was collisionally disrupted.

\textbf{Internal Pressure:} Could internal pressure build-up from sublimated gases cause the nucleus to explode \citep{Samarasinha01}?  The core temperature of the nucleus of A1 is comparable to the Oort cloud equilibrium temperature of just a few degrees above absolute zero.  Heat transport from the surface to the interior by conduction is controlled by the thermal diffusivity, which is proportional to the conductivity and which, in turn, is strongly affected by the particulate nature and porosity of the cometary material.  Laboratory measurements of porous, dielectric powders yield conductivities $\sim$ 10$^2$ to 10$^3$ times smaller than the solid material (Henke et al.~2012).  The expected high porosities and  low thermal diffusivities of cometary material lead to small thermal skin depths that make deep conduction impossible.  Heat applied for a time $\tau$ will conduct over a distance $d \sim (\kappa \tau)^{1/2}$, where $\kappa$ is the diffusivity.  For example, with $\kappa = 10^{-8}$ to 10$^{-9}$ m$^2$ s$^{-1}$,  even in the year between discovery at $r_H$ = 5 au and perihelion at 0.6 au, conducted heat travelled into the nucleus by a characteristic distance only $d \sim$ 0.2 m to 0.5 m.  This distance is so small compared to the  nucleus radius that it is difficult to see how subsurface gas produced by surface heating could have any relevance to the complete disintegration of the nucleus.

\textbf{Rotational Instability:} The remaining possibility for nucleus break-up is also the most plausible.  The timescale for changing the spin angular momentum of a spherical nucleus through outgassing torques is \citep{Jewitt21}

\begin{equation}
\tau_s = \left(\frac{16\pi^2}{15}\right)\left(\frac{\rho_n r_n^4}{k_T V_{th} P} \right)\left(\frac{1}{\dot{M}} \right),
\label{tau_s}
\end{equation}

\noindent where $P$ is the instantaneous spin-period and $k_T$ is the dimensionless moment arm, equal to the fraction of the outflow momentum that exerts a torque on the nucleus.  The median values in a sample of short-period comet nuclei with perihelia in the range 1 $\le q \le$ 2 au are $k_T$ = 0.007 and $P$ = 15 hours (5$\times10^4$ s) \citep{Jewitt21}.  We substitute $\dot{M}$ = 800 kg s$^{-1}$, equal to the sublimation rate at 1 au as measured by Combi's Lyman-$\alpha$ data, on the understanding that this sets a lower bound to the mass loss rate at smaller distances and therefore sets an upper limit to $\tau_s$.  With $\rho_n$ = 500 kg m$^{-3}$, $r_n$ = 600 m, and $V_{th}$ = 500 m s$^{-1}$, substitution into Equation \ref{tau_s} gives $\tau_s < 5\times10^6$ s (0.16 year, or 2 months), which compares to the 6 weeks (0.12 year) spent by A1 with  $r_H <$ 1 au.  While this is not proof that A1 disintegrated through a rotational instability,  given the nominal nucleus parameters and measured mass loss rate, rotational instability does offer a plausible mechanism for nucleus disintegration.  

Rotational breakup is expected to launch fragments  with a velocity dispersion comparable to the tangential speed of the nucleus due to its rotation.  For a strengthless nucleus, this equals the gravitational escape speed from the primary, in this case $\sim$0.3 m s$^{-1}$.  In contrast, the Monte Carlo models show that larger speeds are required to fit the head width of the debris trail.  For example, with $V = V_0 \beta^{1/2}$ and $V_0$ = 550 m s$^{-1}$, millimeter sized particles ($\beta$ = 0.001) would have $V \sim 17$ m s$^{-1}$, about 60 times the escape speed.  We conjecture that these higher speeds result from gas drag acceleration following the exposure and intense sublimation of previously buried ices caused by rotational breakup at $r_H \sim$ 0.8 au. 

Very large particles and boulders would not be substantially accelerated by gas drag and should leave the disintegrating nucleus at about the escape velocity of the primary.  In the $\sim$3 months elapsed between the first signs of breakup and the HST observations, such slow-moving fragments would travel $\sim$2000 km, a distance subtending 1\arcsec~to 2\arcsec~in the plane of the sky (c.f.~Table \ref{geometry}).  Large fragments should therefore be resolvable in the HST data (the resolution is $\sim$0.08\arcsec) but, nevertheless, remain unseen.  This might reflect the continued disintegration of the fragments, again aided by the new exposure to the heat of the Sun of previously buried volatiles.  The breakup process would  then be catastrophic. Smaller fragments produced by breakup of the primary nucleus would have progressively shorter and shorter spin-up times,  owing to their smaller size  (c.f.~Equation \ref{tau_s}) and to the sudden exposure of large areas of previously buried ice which could amplify the moment arm, $k_T$, by orders of magnitude.  The expected result is a runaway fragmentation cascade.

\subsection{Gas Production Resulting from Nucleus Disintegration}
Disintegration of the nucleus must suddenly expose previously buried ices to the heat of the Sun, leading to a burst in the gas production rate caused by sublimation.  Indeed, measurements of the gas production rate  in the mid-December to January period are highly variable, peaking near $Q_{H_2O} = 2.4\times10^{29}$ s$^{-1}$  in radio (Crovisier et al.~2021), Lyman-$\alpha$ (M. Combi, (private communication)), and near-ultraviolet (Jehin et al.~2021, 2022a, 2022b) observations.  At break up, A1 was about $r_H$ = 0.8 AU from the Sun and $\Delta$ = 0.2 AU from the SWAN/SOHO observatory used to take the Lyman-$\alpha$ data.  The latter  has 1\degr~wide pixels, corresponding to about $w \sim 6\times10^5$ km per pixel at the comet and 1.2$\times10^6$ km for the nominal Nyquist (2 pixel) resolution of the data.    With an isothermal blackbody temperature at 0.8 AU $\sim$310 K, the thermal velocity of hydrogen atoms is $V_{th} \sim 2.5 $ km s$^{-1}$.  This, however, is a strong lower limit to the outflow velocity because of photo-electric heating (e.g.~Combi and Delsemme 1980, Combi et al.~2000).  Based on published models, we adopt a hydrogen outflow speed $V_{th} \sim 10 $ km s$^{-1}$ and estimate the residence time for hydrogen atoms within a Nyquist sampled resolution element as $t_r \sim 2w/V_{th} \sim1.2\times10^{5}$ s (about 1.4 days).  This means that the peak  rate inferred from SWAN/SOHO Lyman-$\alpha$ data should be understood as a measure of the  production rate averaged over 1.4 days.

We are interested to see how $Q_{H_2O}$ compares with estimates of the gas production expected from the break up of the nucleus.  To this end, we consider an idealized model  in which the nucleus consists of  particles which are either refractory or ice, and in which the ratio of ice to refractory masses  is $f_{ice}$.  Both refractory and ice particles are assumed to occupy a differential power size distribution (Equation \ref{distribution}).  To render the problem tractable, we make the simplifying assumption that the nucleus disintegrates instantaneously into  power law distributions of ice and refractory particles, each having radii in the range $a_1 \le a \le a_2$.   The icy component then sublimates at the rate $f_s$ [kg m$^{-2}$ s$^{-1}$], which we calculate from energy balance including terms for radiation and sublimation.  

In the residence time $t_r$, an ice surface will sublimate over a layer thickness 

\begin{equation}
a_s = \frac{f_s t_r}{\rho_n},
\label{a_s}
\end{equation}

\noindent where $\rho_n$ is the density of the particle, assumed equal to the bulk density of the nucleus.  All the ice particles with radii $a \le a_s$ will sublimate away, releasing water molecules and, eventually, producing by photodissociation the hydrogen atoms detected using the SWAN instrument.  Ice particles with $a > a_s$ will also partially sublimate in time $t_r$, but their contribution to the gas flux should be small because, for plausible power law distributions (in particular, for $\gamma$ = 3.5 as determined in sections \ref{debris_mass} and \ref{MC}), large particles present a small fraction of the total particle cross-section.

The fraction of the mass contained in ice particles having $a \le a_s$ is given by

\begin{equation}
F = \frac{\int_{a_1}^{a_s}  a^{3-\gamma} da}{\int_{a_1}^{a_2} a^{3-\gamma} da} 
\end{equation}

\noindent which, for $3 < \gamma <$ 4 and $a_s \gg a_1$ and $a_2 \gg a_1$, simplifies to

\begin{equation}
F =  \left(\frac{a_s }{ a_2 }\right)^{4-\gamma}.
\label{F}
\end{equation}

\noindent  The total ice mass in the undisrupted nucleus, assumed to be spherical, is $M_i = (4\pi/3) \rho_n r_n^3  f_{ice}$.   The production rate averaged over time $t_r$ may be written  $\overline{Q_{H_2O}} = F M_i/(t_r \mu m_H)$, where $\mu$ = 18 is the molecular weight of the water molecule and $m_H = 1.67\times10^{-27}$ kg is the mass of the hydrogen atom.  Substitution of Equations \ref{a_s} and \ref{F} into this expression gives

\begin{equation}
\overline{Q_{H_2O}} = \frac{4\pi \rho_n  r_n^3 f_{ice}}{3 t_r \mu m_H} \left(\frac{f_s t_r }{ \rho_n a_2  }\right)^{4-\gamma}.
\label{water}
\end{equation}

\noindent The equilibrium sublimation mass flux calculated for a blackbody water ice sphere at 0.8 AU is $f_s = 1.7\times10^{-4}$ kg m$^{-2}$ s$^{-1}$. The flux could be smaller if the grain albedo is high, or larger if the grain is anisothermal (albeit then sublimating from a smaller fraction of the grain surface).  We set $a_2$ = 60 m, the largest ``particle'' allowed by the Hubble imaging, and $a_1 = 10^{-7}$ m (however, Equation \ref{water} is insensitive to $a_1$ and its value is unimportant provided $a_1 \ll a_s$). The nominal nucleus radius is $r_n$ = 600 m, and the size distribution index is $\gamma$ = 3.5, as deduced above.  Measured cometary ice/refractory ratios, $f_{ice}$, show a wide range of values, from $f_{ice} \sim$  1 in 67P/Churyumov-Gerasimenko (Marschall et al.~2020), to $f_{ice} <$ 0.2 in C/1995 O1 Hale-Bopp (Jewitt and Matthews 1999) and  $f_{ice}$ = 0.03 to 0.1 in 2P/Encke (Reach et al.~2000).  We adopt $f_{ice}$ = 1/4, recognizing that this value is substantially uncertain. 

Substitution into   Equation \ref{water} gives  $\overline{Q_{H_2O}} = 8.8_{-6.2}^{+12.0}\times10^{29}$ s$^{-1}$, where the error bars reflect only the $\pm$200 m uncertainty in the estimated radius of the nucleus.  This is larger than the measured peak water production rate (2$\times10^{29}$ s$^{-1}$) but shows acceptable agreement given the crude nature of the model calculation and the likelihood that the disintegration was in reality spread over a finite period not impulsive, as modeled.    
We conclude that complete disintegration of the nucleus into a power law particle size distribution is consistent both with  the optical brightness of the debris cloud and with the surge in the water production rate measured using Lyman-$\alpha$.  

Future improvements to this model could include a treatment of the initial, optically-thick phase of the expanding disintegration cloud, when self-shielding will suppress and delay the sublimation surge relative to the estimate given here.  Also needed is a treatment of the gas drag interaction with cometary solids in a fully disintegrated body,  responsible for the size-dependent acceleration of refractory particles into the coma and surviving debris field.  Furthermore, several of the parameters needed to accurately model nucleus disintegration remain unmeasured, and most other disintegrating comets are observationally even less-well characterized than A1.  It is obvious, even from these simple considerations that many more detailed observations,  across a wide range of wavelengths and with adequate temporal sampling, will be needed to better understand what is likely to be the dominant destructive cometary process.

\clearpage

\section{SUMMARY}
We present both high resolution and wide field observations of disintegrating long-period comet C/2021 A1 (Leonard) taken to study the nature of its demise.

\begin{itemize}
\item The pre-disintegration radius of the nucleus, estimated using two methods, was $r_n = 0.6\pm0.2$ km.
After breakup, which began in mid-December 2021 and may have continued for weeks, no nucleus fragments larger than about $r_n$ = 0.06 km (i.e.~$< 10^{-3}$ of the primary mass) survived.  
\item The observed debris cloud consists of sub-millimeter and larger particles, with a differential power law size distribution  having index $\gamma$ = 3.4$\pm$0.1 and 3.5$\pm$0.1, as estimated by two different methods.  The observational constraints are consistent with equality between the mass of the debris cloud and the mass of the primary nucleus, indicating a total disintegration.
\item Tidal disruption, sublimation, collisional disruption, and explosion following internal pressure build-up in the nucleus all offer implausible  explanations of the disintegration of C/2021 A1.
\item The spin-up timescale  due to outgassing torques for a 600 m nucleus in the orbit of C/2021 A1 is as short as $\sim$2 months, pointing to rotational instability as the likely cause of the disintegration.  
\item A simple model of the exposure and rapid sublimation of previously buried ice indicates a peak gas production rate ($\overline{Q_{H_2O}} = 9_{-6}^{+12}\times10^{29}$ s$^{-1}$) of the same order as the measured peak value ($\overline{Q_{H_2O}} = 2.4\times10^{29}$ s$^{-1}$).
\end{itemize}

\acknowledgments
We thank Michael Combi for a preview of his SWAN data on C/2021 A1 and the anonymous referee for prompt comments on the manuscript. Based on observations made with the NASA/ESA Hubble Space Telescope, obtained from the data archive at the Space Telescope Science Institute. STScI is operated by the Association of Universities for Research in Astronomy, Inc. under NASA contract NAS 5-26555.  Support for this work was provided by NASA through grant number GO-16929 from the Space Telescope Science Institute, which is operated by auRA, Inc., under NASA contract NAS 5-26555.



{\it Facilities:}  \facility{HST}.

\clearpage


\begin{deluxetable}{lccrrrrccrr}
\tabletypesize{\scriptsize}
\tablecaption{Observing Geometry 
\label{geometry}}
\tablewidth{0pt}
\tablehead{\colhead{UT Date \& Time} &  \colhead{$\nu$\tablenotemark{a}} & \colhead{$r_H$\tablenotemark{b}}  & \colhead{$\Delta$\tablenotemark{c}} & \colhead{$\alpha$\tablenotemark{d}}  & \colhead{$\theta_{- \odot}$\tablenotemark{e}} & \colhead{$\theta_{-V}$\tablenotemark{f}}  & \colhead{$\delta_{\oplus}$\tablenotemark{g}} & \colhead{Tel\tablenotemark{h}} & \colhead{Scale\tablenotemark{i}}   & \colhead{Unc\tablenotemark{j}} }

\startdata
2022 Mar 31 18:14-18:26 & 107.4 & 1.756 & 1.942 & 30.8 & 243.6 & 90.2 & -20.4 & Swan Hill & 1408 & $\pm$1.4  \\
2022 Apr 05 23:35-24:04 & 109.2 & 1.833 & 1.910 & 30.9 & 246.4 & 91.7 & -19.8 & HST & 1385   & $\pm$1.6      \\ 
2022 Apr 06 23:32-23:51  &  109.5   & 1.848 & 1.902 & 30.9 & 246.9 & 92.0 & -19.7& HST & 1379   & $\pm$1.6    \\
2022 Apr 10 19:23-19:53 & 110.7 & 1.903 & 1.875 & 30.7 & 249.0 & 93.2 & -19.1 & HST & 1359 & $\pm$1.7	\\
2022 Jun 7 16:36-17:12 & 123.0& 2.698 & 1.715 & 6.8 & 319.3 & 137.0 & +0.2 & HST 	&   1243 & $\pm$3.7	  \\

\enddata


\tablenotetext{a}{True anomaly, in degrees }
\tablenotetext{b}{Heliocentric distance, in au}
\tablenotetext{c}{Geocentric distance, in au}
\tablenotetext{d}{Phase angle, in degrees }
\tablenotetext{e}{Position angle of projected anti-solar direction, in degrees }
\tablenotetext{f}{Position angle of negative heliocentric velocity vector, in degrees }
\tablenotetext{g}{Angle from orbital plane, in degrees}
\tablenotetext{h}{Telescope}
\tablenotetext{i}{Image scale, km arcsecond$^{-1}$ }
\tablenotetext{j}{ 3$\sigma$ ephemeris uncertainty, arcsecond (from JPL Horizons) }

\end{deluxetable}

\clearpage


%
\clearpage

\clearpage
\clearpage

\clearpage

\clearpage
%

%

\clearpage

\clearpage

\clearpage

\clearpage


\clearpage

\begin{thebibliography}{}

\bibitem[Bessell(1990)]{1990PASP..102.1181B} Bessell, M.~S.\ 1990, \pasp, 102, 1181. doi:10.1086/132749


\bibitem[Bohren \& Huffman(1983)]{Bohren83} Bohren, C.~F. \& Huffman, D.~R.\ 1983, Absorption and scattering of light by small particles. New York: Wiley, 1983

\bibitem[Bottke et al.(2005)]{Bottke05} Bottke, W.~F., Durda, D.~D., Nesvorn{\'y}, D., et al.\ 2005, \icarus, 179, 63. doi:10.1016/j.icarus.2005.05.017

\bibitem[Blaauw et al.(2011)]{Blaauw11} Blaauw, R.~C., Campbell-Brown, M.~D., \& Weryk, R.~J.\ 2011, \mnras, 414, 3322. doi:10.1111/j.1365-2966.2011.18633.x

\bibitem[Combi \& Delsemme(1980)]{1980ApJ...237..633C} Combi, M.~R. \& Delsemme, A.~H.\ 1980, \apj, 237, 633. doi:10.1086/157909

\bibitem[Combi et al.(2000)]{2000Icar..144..191C} Combi, M.~R., Reinard, A.~A., Bertaux, J.-L., et al.\ 2000, \icarus, 144, 191. doi:10.1006/icar.1999.6335


\bibitem[Crovisier et al.(2021)]{Crovisier21} Crovisier, J.,  Biver, N., and  Bockelee-Morvan, D. \ 2021, Central Bureau Electronic Telegram 5087 (2021 December 22)

\bibitem[Finson \& Probstein(1968)]{Finson68} Finson, M.~J. \& Probstein, R.~F.\ 1968, \apj, 154, 327. doi:10.1086/149761

\bibitem[Garcia-Trinanes et al.(2019)]{Garcia19} Garcia-Trinanes, P.,  Luding, S. and Shi, H. \ 2019.  Advanced Powder Technology, 30, 2868-2880


\bibitem[Groussin et al.(2019)]{Groussin19} Groussin, O., Attree, N., Brouet, Y., et al.\ 2019, \ssr, 215, 29. doi:10.1007/s11214-019-0594-x

\bibitem[Henke et al.(2012)]{2012A&A...537A..45H} Henke, S., Gail, H.-P., Trieloff, M., et al.\ 2012, \aap, 537, A45. doi:10.1051/0004-6361/201117177

\bibitem[Holmberg et al.(2006)]{2006MNRAS.367..449H} Holmberg, J., Flynn, C., \& Portinari, L.\ 2006, \mnras, 367, 449. doi:10.1111/j.1365-2966.2005.09832.x


\bibitem[Ishiguro et al.(2007)]{Ishiguro07} Ishiguro, M., Sarugaku, Y., Ueno, M., et al.\ 2007, \icarus, 189, 169. doi:10.1016/j.icarus.2007.01.003

\bibitem[Jehin et al.(2021)]{Jehin21} Jehin, E., Moulane, Y., \& Manfroid, J.\ 2021, The Astronomer's Telegram, 15128


\bibitem[Jehin et al.(2022)]{Jehin22} Jehin, E., Moulane, Y., Manfroid, J., et al.\ 2022a, The Astronomer's Telegram, 15186

\bibitem[Jehin et al.(2022)]{Jehinburst} Jehin, E., Moulane, Y., Manfroid, J., et al.\ 2022b, The Astronomer's Telegram, 15189

\bibitem[Jewitt \& Matthews(1999)]{1999AJ....117.1056J} Jewitt, D. \& Matthews, H.\ 1999, \aj, 117, 1056. doi:10.1086/300743

\bibitem[Jewitt \& Luu(2019)]{2019ApJ...883L..28J} Jewitt, D. \& Luu, J.\ 2019, \apjl, 883, L28. doi:10.3847/2041-8213/ab4135

\bibitem[Jewitt et al.(2020)]{2020ApJ...896L..39J} Jewitt, D., Kim, Y., Mutchler, M., et al.\ 2020, \apjl, 896, L39. doi:10.3847/2041-8213/ab99cb

\bibitem[Jewitt(2021)]{Jewitt21} Jewitt, D.\ 2021, \aj, 161, 261. doi:10.3847/1538-3881/abf09c

\bibitem[Jewitt(2022)]{2022AJ....164..158J} Jewitt, D.\ 2022, \aj, 164, 158. doi:10.3847/1538-3881/ac886d

\bibitem[Kim et al.(2017)]{Kim17} Kim, Y., Ishiguro, M., Michikami, T., et al.\ 2017, \aj, 153, 228. doi:10.3847/1538-3881/aa69bb

\bibitem[Knight et al.(2010)]{Knight10} Knight, M.~M., A'Hearn, M.~F., Biesecker, D.~A., et al.\ 2010, \aj, 139, 926. doi:10.1088/0004-6256/139/3/926


\bibitem[Leonard et al.(2021)]{Leonard21} Leonard, G.~J., Aschi, S., Pettarin, E., et al.\ 2021, Minor Planet Electronic Circulars, 2021-A99

\bibitem[Li \& Jewitt(2015)]{2015AJ....149..133L} Li, J. \& Jewitt, D.\ 2015, \aj, 149, 133. doi:10.1088/0004-6256/149/4/133


\bibitem[Marschall et al.(2020)]{2020FrP.....8..227M} Marschall, R., Markkanen, J., Gerig, S.-B., et al.\ 2020, Frontiers in Physics, 8, 227. doi:10.3389/fphy.2020.00227


\bibitem[Pansecchi \& Fulle(1990)]{1990A&A...239..369P} Pansecchi, L. \& Fulle, M.\ 1990, \aap, 239, 369

\bibitem[Reach et al.(2000)]{2000Icar..148...80R} Reach, W.~T., Sykes, M.~V., Lien, D., et al.\ 2000, \icarus, 148, 80. doi:10.1006/icar.2000.6478


\bibitem[Samarasinha(2001)]{Samarasinha01} Samarasinha, N.~H.\ 2001, \icarus, 154, 540. doi:10.1006/icar.2001.6685





\bibitem[Wolf et al.(2018)]{2018PASA...35...10W} Wolf, C., Onken, C.~A., Luvaul, L.~C., et al.\ 2018, \pasa, 35, e010. doi:10.1017/pasa.2018.5

\bibitem[Zhang et al.(2021)]{Zhang21} Zhang, Q., Ye, Q., Vissapragada, S., et al.\ 2021, \aj, 162, 194. doi:10.3847/1538-3881/ac19ba

\bibitem[Zubko et al.(2017)]{Zubko17} Zubko, E., Videen, G. Shkuratov, Y., et al.\ 2017, JQSRT, 202, 104. doi:10.1016/j.jqsrt.2017.07.026


\end{thebibliography}
\end{document}